\documentclass {article} 
\usepackage{arxiv}
\usepackage{booktabs} 
\usepackage{algorithm}%
\usepackage{algpseudocode}%
\usepackage{hyperref}
\usepackage{amsmath}
\usepackage{xcolor}
\usepackage{xspace}
\usepackage{soul}
\newcommand{\ignore}[1]{}
\usepackage[textsize=tiny]{todonotes}
\usepackage{graphicx}
\usepackage{boldline}
\usepackage[export]{adjustbox}
\usepackage[caption=false]{subfig}
\usepackage[font=small]{caption}
\usepackage{adjustbox}
\usepackage{enumerate}
\usepackage{amsfonts}
\usepackage{colortbl}
\usepackage{float}
\usepackage{rotating}
\usepackage{setspace}
\usepackage{gensymb}	
\usepackage[official]{eurosym} 
\usepackage[flushleft]{threeparttable}
\usepackage{enumitem}
\usepackage{array}
\title{Performance investigation on the different wave energy converters layouts in real wave scenarios}

\author{
Erfan Amini\\
Coastal and offshore structures engineering group\\
School of Civil Engineering\\
	The University of Tehran\\
	 Iran\\
	 \texttt{erfan.amini@ut.ac.ir} \\
\And
  Danial Golbaz \\
  Coastal and offshore structures engineering group\\
  School of Civil Engineering\\
  The University of Tehran\\
   Iran \\
  \texttt{Dgolbaz@ut.ac.ir} \\
  \And
  Fereidoun Amini \\
   School of Civil Engineering \\
  Iran University of Science and Technology\\
   Iran \\
  \texttt{Famini@iust.ac.ir} \\
  }

\begin{document}

\maketitle

\begin{abstract}
Wave energy is a broadly accessible renewable energy source, but principally it is not unexploited still and fully developed.  Wave energy technologies need more investigations to reach commercial developments. In this paper, an array of interacting wave energy converters (WECs) with various arrangements and layout size is considered. The WEC model is a symmetrical, spherical and fully submerged buoy with three tethers. A frequency-domain model is derived by using linear potential theory. Numerical simulations and parametric examinations are designed in order to evaluate the distance influence among buoys with pre-defined arrangement  on the total power output of the array, based on the  irregular wave models. With regard to validating the best-found layout, we perform a landscape analysis experiment using a grid search technique.  These experiments are evaluated in four real wave scenarios from south of Australia. Experiments show that the distance between WECs and also relative angle of the layout to significant wave direction plays a pivotal role to harness more power from the incident waves. Furthermore, we observe that increasing the number of WEC in the layout leads to a rise in the optimal distance among WECs. The maximum power of an array produced by a 5 buoy layout arrangement with 185 $m$ distance among buoys in the Tasmania wave scenario with relative angle of 63 degree to significant in-site wave direction . 

\end{abstract}
\doublespacing

\sloppy

\section{Introduction}
Wave energy is expected to provide towards the development to carbon-free electricity generation. The theoretical computation of wave energy potential over the oceans is projected to be in the order of 1–10 TW~\cite{barstow2008wave}, which can cover the current global energy demand~\cite{energy2016international}. This tremendous
potential has dragged the consideration of the research societies, which has proved that harnessing electric power from ocean waves is possible~\cite{clement2002wave,antonio2010wave}. However, wave energy is still considered high-priced in comparison with wind or solar renewable energy. This challenge makes a  technological obstacle for researchers to develop this technology, so the commercialisation progress of wave energy technology has been advancing gradually.

Wave energy converters (WEC) are planned to be stationed in a wave farm constituted of many generators similar to wind turbines. Each generator (converter) interacts with other ones, by harnessing, radiating and diffracting the incident waves. The position of converters in the array which is scattered through the farm has a direct relationship with the performance of the farm because hydrodynamic interactions between them can be constructive or destructive. These interactions are depending on the configuration of the array. 
Consequently, it is the main reason to investigate these interactions in order to apply them to reinforce the total power output. There are many relevant publications with this subject by several R\&D units across Europe in the past by the pioneering works~\cite{budal1977theory,evans1976theory,budal1980interacting, thomas1981arrays,simon1982multiple},  and it is still an interesting research field, as several investigations
have been published recently~\cite{castro2020design,neshat2020hybrid,bonovas2020modelling,calvario2020oil} . Table~\ref{tab:survey} demonstrates a  briefly survey some of the recent literature in the various aspects of WECs optimisation. 

\begin{table}[htb]
    \small
    \caption{A briefly survey some of the recent literature on the layout, PTO parameters and design optimisation of wave energy converters }
    \label{tab:survey}
    \centering
    \begin{tabular}{c|c|c|c|c|c}
        \toprule
       Objective &  WECs type & WECs Number & Method & Year & reference   \\
      \midrule
        Design \& PTOs& submerged & 2  & Experimental observations  & 2020 & ~\cite{castro2020design}     \\ \hline
        Layout \& PTOs & fully-submerged  & 4, 16  &  Cooperative EAs & 2020 &~\cite{neshat2020hybrid}      \\ \hline
        Design \& PTOs& fully-submerged  & 1  & Hybrid EAs  & 2020 & ~\cite{sergiienko2020design}     \\\hline
        Layout& fully-submerged  & 50, 100   & Multi-strategy EAs  & 2020 &  ~\cite{10.1145/3377930.3390235}    \\\hline
       Design \& PTOs &  heaving WEC &  1 & Evolutionary and GA  & 2020 &~\cite{bonovas2020modelling}     \\\hline
       PTOs   & oscillating wave surge converter  &  1 &  GA & 2020 & ~\cite{calvario2020oil,jusoh2020parameters}     \\ \hline
         Design   & sloped-motion WEC  & 1  & Heuristic optimization  & 2020 &~\cite{rodriguez2020hydrodynamic}      \\ \hline
        PTOs& oscillating water column–based  &  1 & Water cycle algorithm  & 2020 &~\cite{m2020water}      \\\hline
       PTOs &hinged-type WECs   &  1 &  Experimental observations & 2020 & ~\cite{chen2020hydrodynamic}     \\\hline
       PTOs & oscillating wave surge converter   & 1  & GA and ML  & 2020 & ~\cite{liu2020prediction}     \\\hline
         Layout &  submerged & 25  &  PSO & 2020 & ~\cite{izquierdo2020layout}     \\ \hline
          Design  & submerged flat plate  & 1  & GA  & 2019 &~\cite{esmaeilzadeh2019shape}      \\ \hline
       Design \& Layout  & cylindrical heaving WECs  & 3, 5, 7   & GA  & 2019 & ~\cite{lyu2019optimization}     \\\hline
       Design &  submerged & 2  &  GA &  2019& ~\cite{kelly2019shape}     \\\hline
       Layout & fully-submerged   & 4, 16  &  Smart heuristic & 2019 & ~\cite{neshat2019new}     \\\hline
        Layout & fully-submerged  & 4, 16  & Nuro-adaptive EA  & 2019 & ~\cite{neshat2019adaptive}     \\\hline
        PTOs & freely floating  & 2   & EAs  & 2019 & ~\cite{jabrali2019viscous}     \\\hline
       Design & hinge-barge WEC  & 2  &  gradient-based method  & 2019 &~\cite{wang2019geometric}      \\\hline
         Design & fully-submerged  & 1, 2, 3  & GA, PSO  & 2019 & ~\cite{faraggiana2019design}     \\ \hline
         Layout \& PTOs   & fully-submerged  & 16  & Hybrid EAs  & 2019 & ~\cite{10.1145/3321707.3321806}     \\ \hline
       Layout \& PTOs & fully-submerged  &  4, 9 & Heuristics  &2019  & ~\cite{vatchavayi2019heuristic}     \\\hline
       Layout & heaving WEC  & 1  & GWO  & 2019 & ~\cite{amini2019investigating}     \\\hline
      Layout  & heave-constrained cylinder  & 5  & improved GA  & 2018 &~\cite{sharp2018wave}      \\\hline
        Layout & fully-submerged  &  4, 16  & Local search & 2018& ~\cite{neshat2018detailed}     \\\hline
       Layout &  oscillating WEC & 3, 5, 8  & improved DE  & 2018  & ~\cite{fang2018optimization}     \\\hline
       Layout \&LCoE & fully-submerged  & 4, 9, 36  &  Multi-objective EAs & 2018 & ~\cite{arbones2018sparse}     \\\hline
       PTOs   &  submerged & 1  & Hidden GA  &2018  & ~\cite{abdelkhalik2018optimization}     \\ \hline
       Layout \& PTOs & submerged  & 4, 7, 9, 14  & hybrid GA  &2018  & ~\cite{giassi2018layout}     \\ \hline
        
       \midrule

    \end{tabular}
\end{table}
This paper concentrates particularly on the arrangement optimisation of WEC arrays and shows the effectiveness of the inter-distance among WECs to produce more power. In order to establish a wave farm, an optimal layout is chosen to maximise the power conversion; however,  the number of WECs is a significant factor. We evaluate a various number of WECs as an array, arrangements and separations and report the performance of the layouts using both q-factor, power of each converter and total power output. The distances between the WECs, and the wave farm size are constrained, which is a more realistic approach for the study of WEC arrays. Finally, a landscape analysis experiment is performed with regard to evaluate the position effect of each WEC in the array's power output using a grid search approach.

This paper is structured into five sections. Section 2 presents a brief description of the hydrodynamic WEC array interaction model, modelling the wave climate and the equations used to compute the produced power. Section 3 expresses the layout assessment routine and presents the strategy to explore the optimal position of the WECs in the farm.  Section 4 discusses the array layout investigation results in terms of performance and optimal array layout solutions. Subsequently, Section 5 summarises the principal finding of the paper.



\section{Numerical Modelling}\label{Section2}

\subsection{Wave energy converter}
\begin{figure}[htb]
    \centering
    \includegraphics[width=0.6\linewidth]{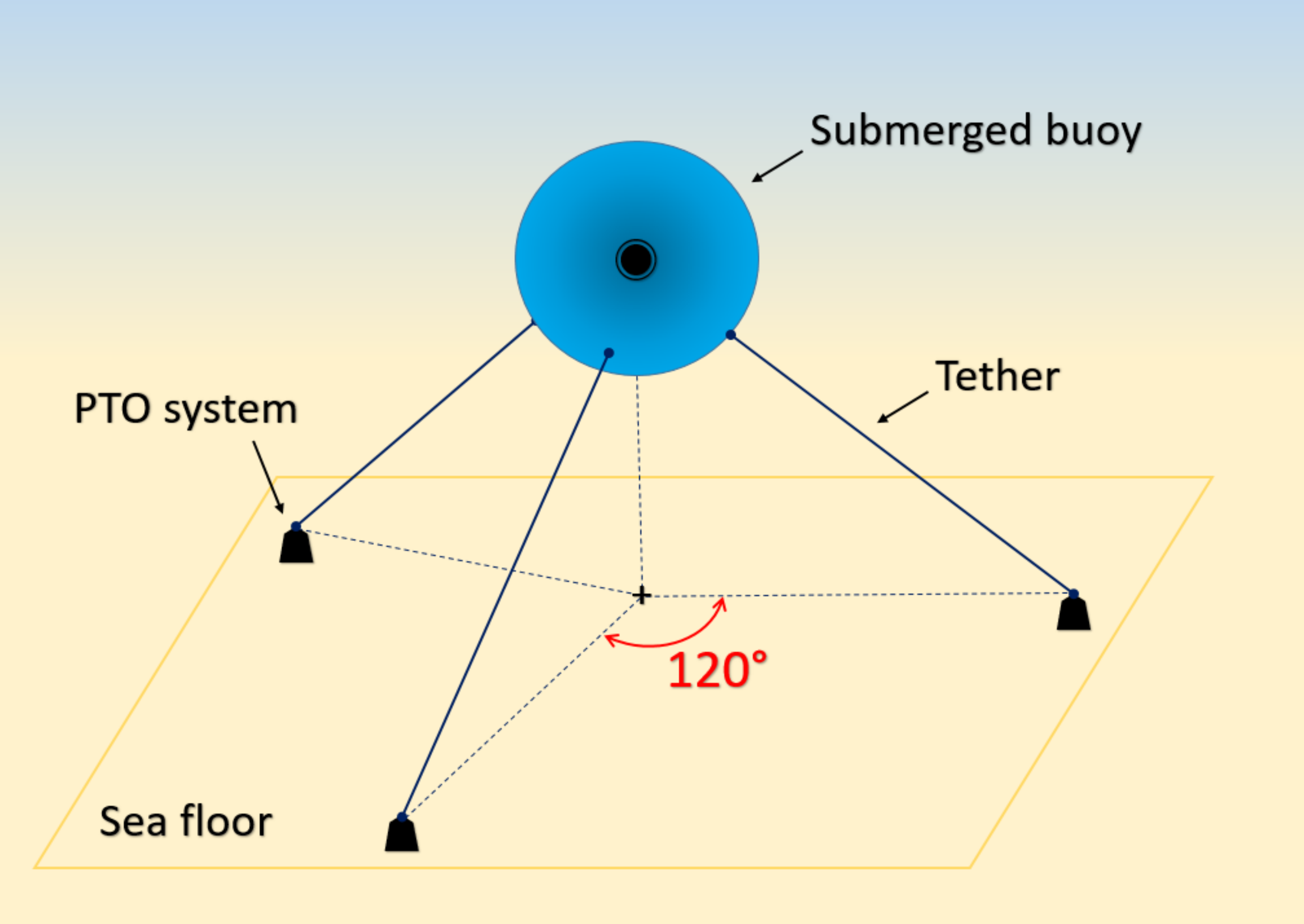}
    \caption{Schematic representation of modelled point absorber wave energy converter (adapted from~\cite{Sergiienko2016} ).}
    \label{fig:schem}
\end{figure}

In this study, a wave energy converter with three tethers mooring system is considered which has a fully submerged spherical buoy attached to seabed by these tethers which is shown in figure~\ref{fig:schem}.
This model is developed in MATLAB and modified in 2020~\cite{matlab_model_serg_2020}. The WEC details are:  buoy radius=5 ($m$) ,  submergence depth=3 ($m$), water depth=50 ($m$), buoy mass=376 (tonnes), buoy volume=523.6 ($m^3$), tether angle=55 (degree), PTO stiffness=$2.7*10^5 (N/m)$, PTO damping= $1.3*10^5 (Ns/m)$. \\
This buoy is floating at sea moves in six degrees of motion, however three degrees (surge, heave, sway) of freedom affect on motions of the WECs. Based on these degrees, motion equation can be written on frequency domain.
\begin{equation}\label{motion.eq}
\begin{split}
\Sigma F & = m\ddot{z},\\
&= F_m + F_{hs}+W+F_R + F_{PTO}+F_{W_k}+F_{VD}
\end{split}
\end{equation}
where $F_m$ is the mooring force, $F_{hs}$ is the hydro-static force resulting from buoyancy, $W$ is the body weight, $F_R$ represents the hydrodynamic forces that include added mass and wave damping forces, force resulted by PTO system is $F_{PTO}$, $F_{W_k}$ represents the vertical components of the wave exciting force and $F_{VD}$ is the vertical viscous drag force~\cite{motionequation}.
This equation -which is also known as Cummins equation- used in order to describe a time-domain response of the WECs in waves, can be utilized and rewritten as:

\begin{equation} \label{Cummins B1}
(m+A_\infty)\ddot{z}+ \int_{0}^{t} K_{rad}(t-\tau)\dot{z}(\tau)\mathrm{d}\tau+ Cz= F_{exc}+F_{pto} +F_{hs} 
\end{equation}
where $m$ is a buoy mass, $A_{\infty}$ is the infinite-frequency added mass coefficient, $C$ is the hydro-static stiffness, $K_{rad}(t)$ is the radiation impulse response function, $F_{exc}$ is the wave excitation force, $F_{pto}$ is the load force exerted on the buoy from the power take-off system~\cite{Cummins1962}.
Free surface elevation height is resulted from a linear superposition which consists of some wave characteristics, in irregular waves. This is usually determined by a wave spectrum which describes distribution of energy in vast number of wave frequencies. Significant wave height and peak period are utilized as the basic identification of wave in this spectrum. Excitation force can be calculated as a real part in integral over the total wave frequencies .

\begin{equation}\label{Fexc}
F_{exc}=\textbf{R} \left[ R_f \int_{0}^{\infty} \sqrt{2S(\omega_r)} F_x \  e^{i(\omega_r t+\phi)}   \mathrm{d}\omega_r \right] = \int_{-\infty}^{+\infty}\eta_{(\tau)}f_e (t-\tau)\mathrm{d}_{(\tau)}
\end{equation}
Where \textbf{R} denotes the real part of the equation, $R_f$ is the ramp function, $F_x$ is the excitation vector consists of amplitude and phase of the wave, $S$  is the wave spectrum, $\phi$  is the stochastic phase angle, $\eta_{\tau}$  represents water elevation and $f_e$  is the element of force vector ~\cite{amini2019investigating}.
The load force of PTO is modelled as a linear spring-damper system. However, the system is modelled by a repulsive energy potential in order to constrain the motions of the buoy.

\begin{equation}\label{pto}
F_{pto}=-B_{pto}\dot{z} - K_{pto}z
\end{equation}
\begin{align}\label{fhs}
F_{hs}=& - K_{hs,\text{min}}(z-z_{\text{min}})u(z_{\text{min}} - z) \notag \\
& -K_{hs,\text{max}}(z-z_{\text{max}})u (z-z_{\text{max}})
\end{align}
Where in equation \ref{pto} $K_{pto}$ and $B_{pto}$ are control parameters which represent stiffness and damping of PTO and in equation \ref{fhs}  $u$ is Heaviside step function, $K_{hs,\text{min}}$ and  $K_{hs,\text{max}}$ are the hard stop spring coefficients, $z_{\text{max}}$ as well as $z_{\text{max}}$ are the stroke limits which are related to the nominal position of the converter. It is important to note that for computing useful absorbed energy, the effect of this force is not considered~\cite{sergiienko2017performance}.

In order to calculate the produced energy by each buoy, sum of three forces is necessary: wave excitation$(F_{exc,p}(t)) $, force of radiation $(F_{rad,p}(t))$,  power take off force $ (F_{pto,p}(t))$. All interaction forces are considered in $F_{exc}$ whether they are destructive or constructive. Plus that, control parameters of PTO which are used for each mooring line, and also hydrodynamic parameters derived from semi-analytical model are taken in order to compute the total power output of a buoy array.\\
To calculating average power absorbed by the wave farm, several variables have to take in to account as follows.
\begin{equation}\label{av-powerp}
P_n (H_s,T_p)= \int_{0}^{2\pi}\int_{0}^{\infty} 2S_n (\omega)D(\beta)p(\beta,\omega)\mathrm{d}\omega \mathrm{d}\beta
\end{equation}
Where $P_n (H_s,T_p)$ is the average power absorbed by the wave farm in a regular wave of unit amplitude, $S_n (\omega)$ is the irregular wave spectrum which is calculated with the Bretschneider spectrum and $D(\beta) $ represents the directional spreading spectrum particularly for this site which is come from directional wave rose ~\cite{flavia2017pn-hs-tp}. $\omega$ is the wave frequency and $p(\beta,\omega)$ is the wave angle which are based on the $z(\beta,\omega)$ ~\cite{neshat2019new}, which can be calculated by equation \eqref{Cummins B1} at the beginning of this section. 
The wave farm at a certain test site is generated by total mean annual power $P_{array}$, and for calculating that, the contribution of energy absorption from a wave climate in each state should be summed up:

\begin{equation}
P_{array} =\sum_{n=1}^{N_s} O_n (H_s,T_p)P_n(H_s,T_p)
\end{equation}

where $N_s$ is a number of chosen sea state, $H_s$ is the significant wave height and $T_p$ is the peak wave period for each sea state, $O_n(H_s,T_P)$ represents the probability of occurrence of sea state which is stemmed from the wave scatter diagram and $P_n(H_s,T_p)$ is a power which wave farm produces in the $n $th sea state   \cite{neshat2019new}. Significant wave height and peak wave period are statistics of a sea state which can be referred to the condition of the ocean/sea surface. To calculate    $P_n(H_s,T_p)$ in irregular waves, it is needed to sum all power contribution in each frequency and wave direction.

It is important to measure the effectiveness of interactions between converters. Due to this, q-factor is defined to play as a determinant, where if $ q>1$, then it has positive effect on the total energy of an array, otherwise, the interactions are destructive.
\begin{equation}\label{q-factor}
q=\frac{P_{array}}{NP_{isolated}}
\end{equation}
where $P_{isolated}$ is the power that an isolated WEC generates,  $N$ is the number of converters \cite{neshat2019adaptive}.

\subsection{Wave Resource}
According to the previous works~\cite{neshat2019new, 10.1145/3321707.3321806}, four different sea sites were chosen for this study. 
  Significant wave height and scattered diagram for one of the chosen locations can be seen in figure \label{fig:sydney}, which implies to distinctions in data and various scenarios in each location. It is evident from the scattered diagram that peak wave period are mostly from 6 to 11 seconds, and the height of them are between 1 and 3 meters. The direction of waves are also can be interpreted from this directional wave rose diagram.

\begin{figure}[htb]
    \centering
    \includegraphics[width=\linewidth]{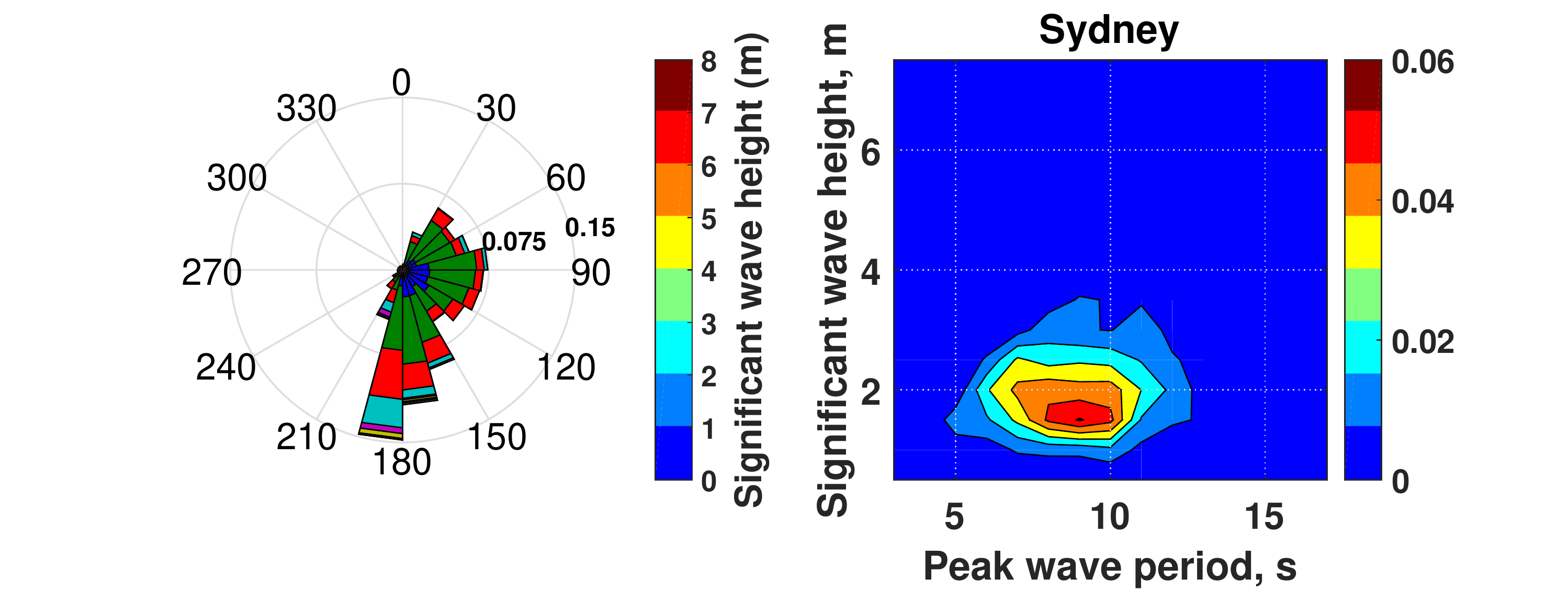}
    \caption{The wave climate at the Sydney deployment site located in Australia's east coast from~\cite{neshat2019new}}
    \label{fig:sydney}
\end{figure}

Limitations are inevitable while an array is about to design. Each farm is constrained by maximum area and minimum distance between WECs. Firstly, minimum separation between buoys $(R')$ has to be 50 meters to provide a safe pass for vessels. Secondly, although area grows by increasing the number of buoys, it has to be constrained within the area $\Omega$, where $
\Omega=l \times \omega ,\quad l=\omega= \sqrt{N\times 20000} \ \text{m}
$ ~\cite{10.1145/3321707.3321806}.  
The Bretschneider spectrum is used for modelling irregular waves in this study. This spectrum is called a modified, two parameters Pierson Moskowitz spectrum which is based on significant wave height and peak period. These two are highly depend on wind speed and its direction. Moreover, fetch and location of storm have considerable effects on the Bretschneider spectrum too ~\cite{zwolan2012bretchneider}. 

\begin{subequations}
\begin{align}\label{Brets}
S(f)= & \tfrac{H_{m0}^2}{4}(1.057f_p)^4 f^{-5} \text{exp}\left[-\frac54 (\tfrac{f_p}{f})^4\right] \\
 & A= \tfrac{H_{m0}^2}{4} (1.057 f_p)^4 \approx \tfrac{5}{16}H_{m0}^2 f_p^4 ,\\
 & B=(1.057 f_p)^4 \approx \tfrac54 f_p^4 
\end{align}
\end{subequations}

\subsection{Annual Energy Output }
The optimal designs of wave farm for four different locations in Australia use power matrices of various configurations (i.e. different layout geometry, WEC distance and wave direction). To be more precise, the goal is to figure out the best wave farm layout configuration with optimal separation between WECs and rotation angle,  the one that are providing the highest annual energy output (AEO), at each research site. For this aim, based on the number of WECs, different layouts can be deployed with various orientations and separation between converters. Note that, there are certain limits for distances, hence it depends on the number of WECs. Similarly, the number of converters in a farm allows configurations to be chosen. In this work, as number of converters increases, the directions of wave propagation witness downward trend. The below equation enable calculation of AEO with considering the number of converters, variety of wave directions and allowable distance with 1 meter interval.
\begin{equation}\label{AEO}
\textbf{meanAEO}_{\text{(each wave scenario)}}=\sum_{i=0}^{max \alpha}\sum_{j=50}^{l} \text{P}_{{array}_{(i,j)}}
\end{equation}
Where $\alpha$ is the direction of wave with 10 degree interval, except when $N$ is 5, the interval changes to 9 degrees ranging from 0-63 degrees.

\section{Layout Assessment Routine}
According to the mentioned equations in section \ref{Section2}, following outcomes is obtained. Four different layout is considered regarding to the number of buoys, which their details is thoroughly described.

\subsection{Two-buoy Layout}
\begin{figure}[htb]
    \centering
    \includegraphics[width=0.5\linewidth]{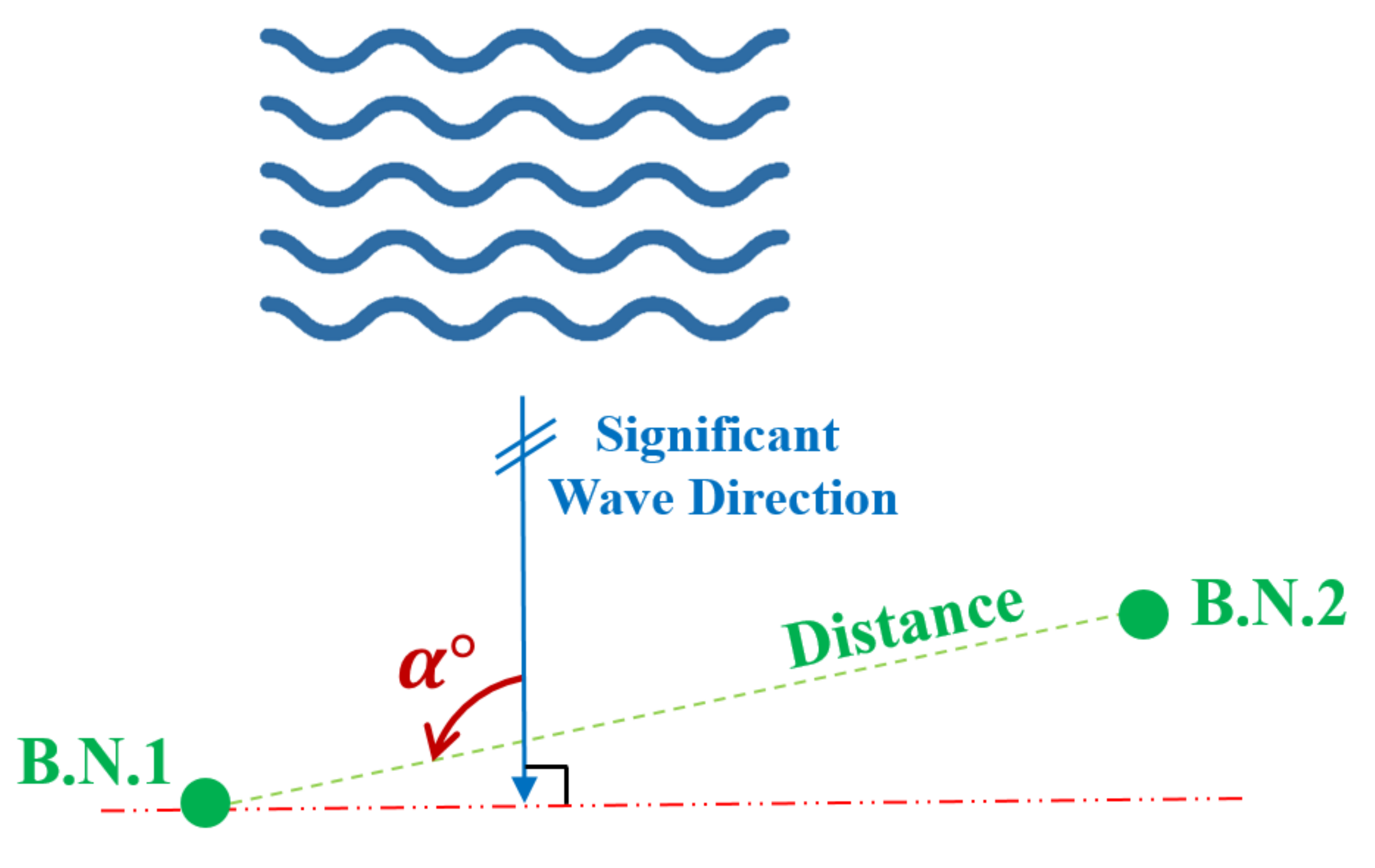}
    \caption{Two buoy layout regard to significant wave direction.}
    \label{fig:BN2}
\end{figure}

There are two buoys in this array which there would be one line connecting them to each other. Significant wave direction is obtained by considering one-third of maximum waves in scattered diagram and it would be 232.5 degree. The angle between significant wave direction and hypothetical line is considered to be alpha which is clearly illustrate in figure~\ref{fig:BN2}. The interval of alpha is chosen to be tested in every 10 degree, therefore there would be eighteen different alpha ranging from 0 to 170 degree. Consideration of this range is because of preventing to calculate results that is already calculated. Likewise, distances is also assumed to change per meter between the allowable period. At the end, three measurements which are power of each buoy, farm power and q factor are taken in each step.

\subsection{Three-buoy Layout}

\begin{figure}[htb]
    \centering
    \includegraphics[width=0.5\linewidth]{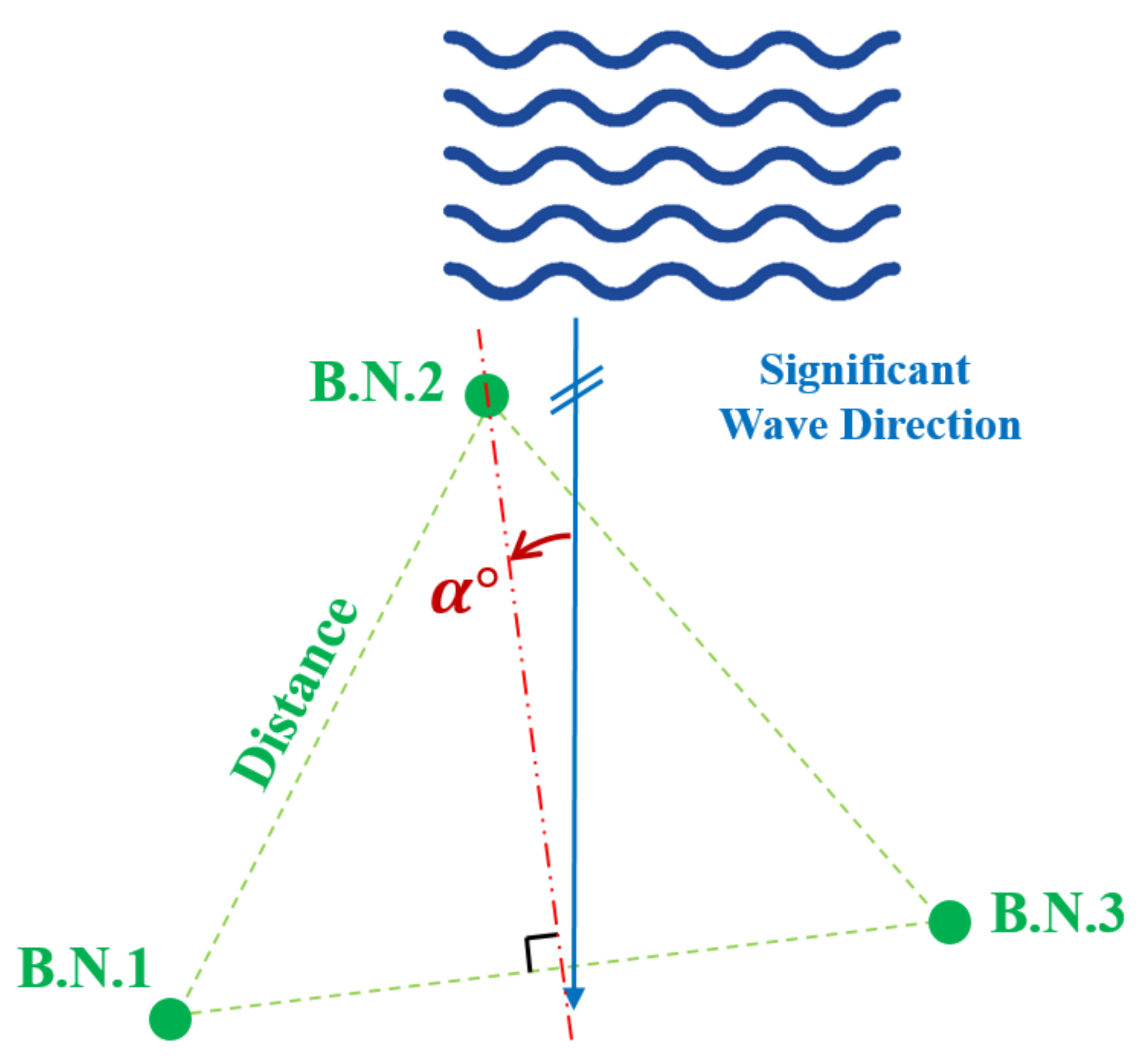}
    \caption{Three buoy layout regard to significant wave direction.}
    \label{fig:BN3}
\end{figure}

When there are three buoys to consider in an array, one of the most common geometry is equilateral triangle. If some lines are used to connect these buoys to each other, angles between them will be 60 degree that makes one of the converters as the vertex of triangle. There is a line from this converter perpendicular to the line which is connecting two other buoys. The angle between significant wave direction which is 232.5 degree and this perpendicular line is alpha which is shown in ~figure \ref{fig:BN3}, and this parameter has twelve degrees from 0 to 110 which changes per 10 degree.  Note that for every chosen alpha, distance is changing per meter to cover all area, and in each step power of buoy, array power and q factor is calculated.

\subsection{Four-buoy Layout}

\begin{figure}[htb]
    \centering
    \includegraphics[width=0.5\linewidth]{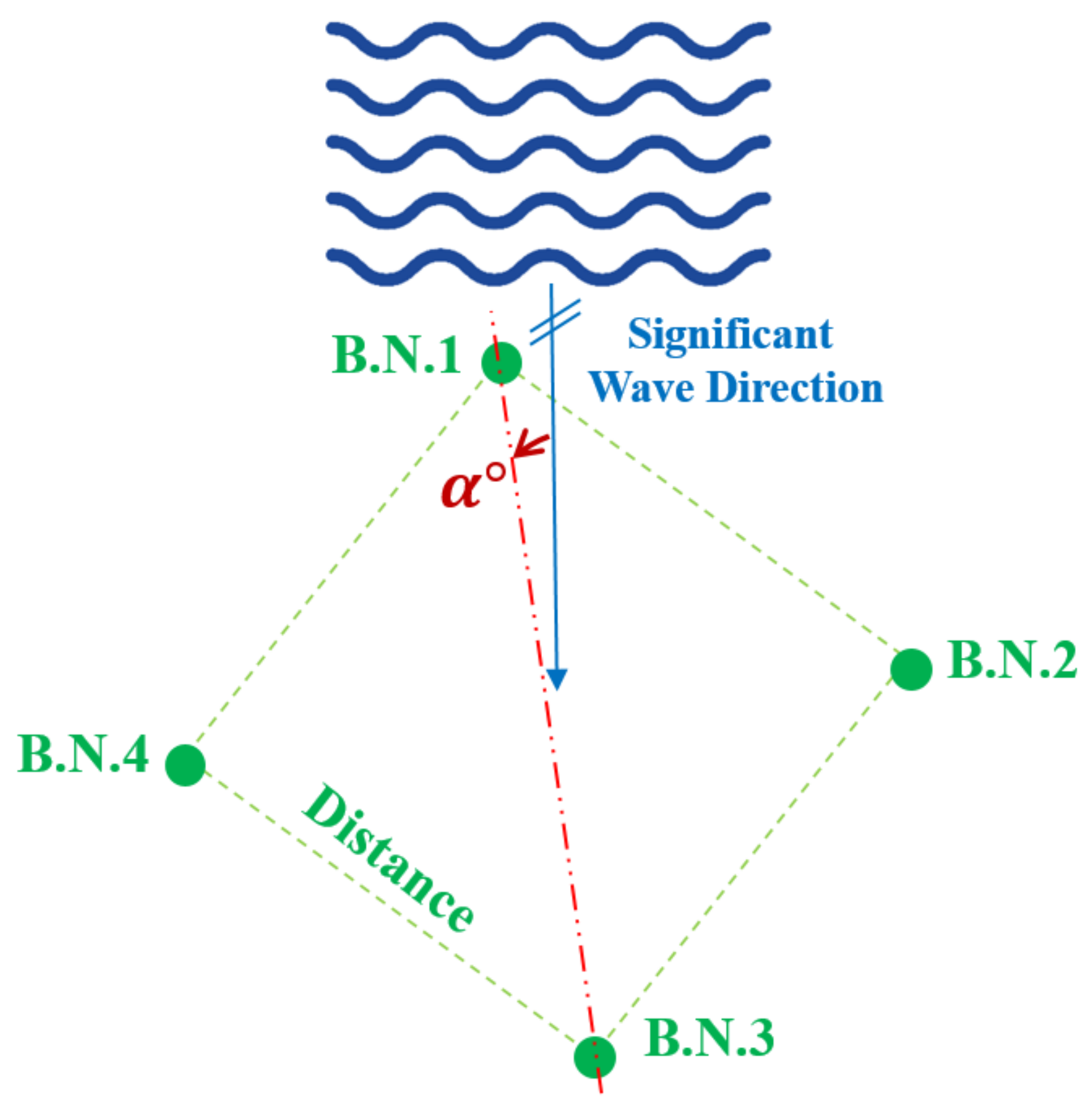}
    \caption{Four buoy layout regard to significant wave direction.}
    \label{fig:BN4}
\end{figure}

To choose a configuration for four buoys, a regular quadrilateral is taken into account. To describe alpha in this layout, firstly significant wave direction is needed to be determined, which it is 232.5 degree. Secondly, a hypothetical line from one converter to the furthest buoy should be drawn. For example, if converters are numbered clockwise and the closest buoy to the front wave is buoy number one, the line should be drawn from 1 to 3, exactly like figure ~\ref{fig:BN4}. Finally, the angle between this line and significant wave direction is alpha, and the range of this is from 0 to 80 degree which	has	9 different amounts with equal  intervals. Similarly, distances interval is 1 meter during each alpha. Once again, some measurements such as farm power, power of each buoy and q factor are computed and stored per distance.

\subsection{Five-buoy Layout}

\begin{figure}[htb]
    \centering
    \includegraphics[width=0.5\linewidth]{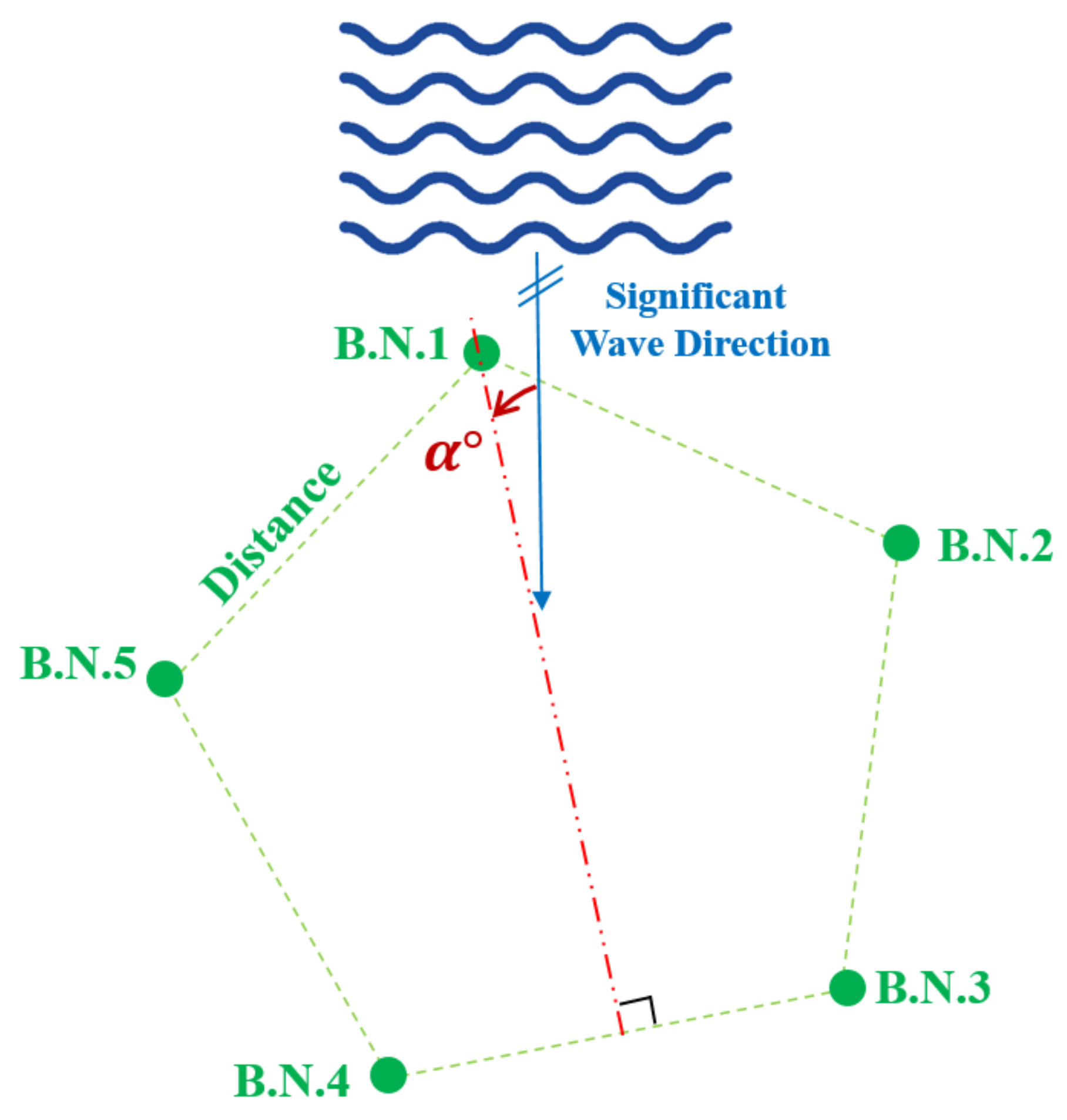}
    \caption{Five buoy layout regard to significant wave direction.}
    \label{fig:BN5}
\end{figure}

In this layout,	five similar converters form a farm in shape of a regular pentagon. Significant wave direction in this layout is 172.5 degree which is illustrated in figure \ref{fig:BN5} with blue arrow. Converters are numbered clockwise and the first number is started from the closest buoy to the front wave. As it can be seen in figure \ref{fig:BN5}, each converters have the longest distance with two buoys, in this case the furthest converters to buoy number 1 are number 3 and 4. If a perpendicular line is drawn from first converter to the connecting line between furthest converters, the angle between significant wave direction and the perpendicular line represents alpha. The range for alpha is from 0 to 63 with 9 degree interval, so there are 8 alphas to test in this layout. It is needed to say that interval of distances in each alpha is 1 meter and for every distance power of buoy farm power and q factor is calculated. The details of all results are discussed comprehensively in section \ref{section5}. 

\section{Results and Discussions}\label{section5}
This section represents the results of different array layouts when the number of buoys changes from 2 to 5 in four considered locations in Australian coasts. The results indicate sensitivity analysis of the q factor and farm power due to the wave direction ($\alpha$), and distance between WECs, for each layout. It is worth to mention that there are sixteen conditions on this study which will be discussed in detail.

\subsection{Sensitivity of two-buoy array performance to distance}
\begin{figure}[htb]
    \centering
    \includegraphics[width=\linewidth]{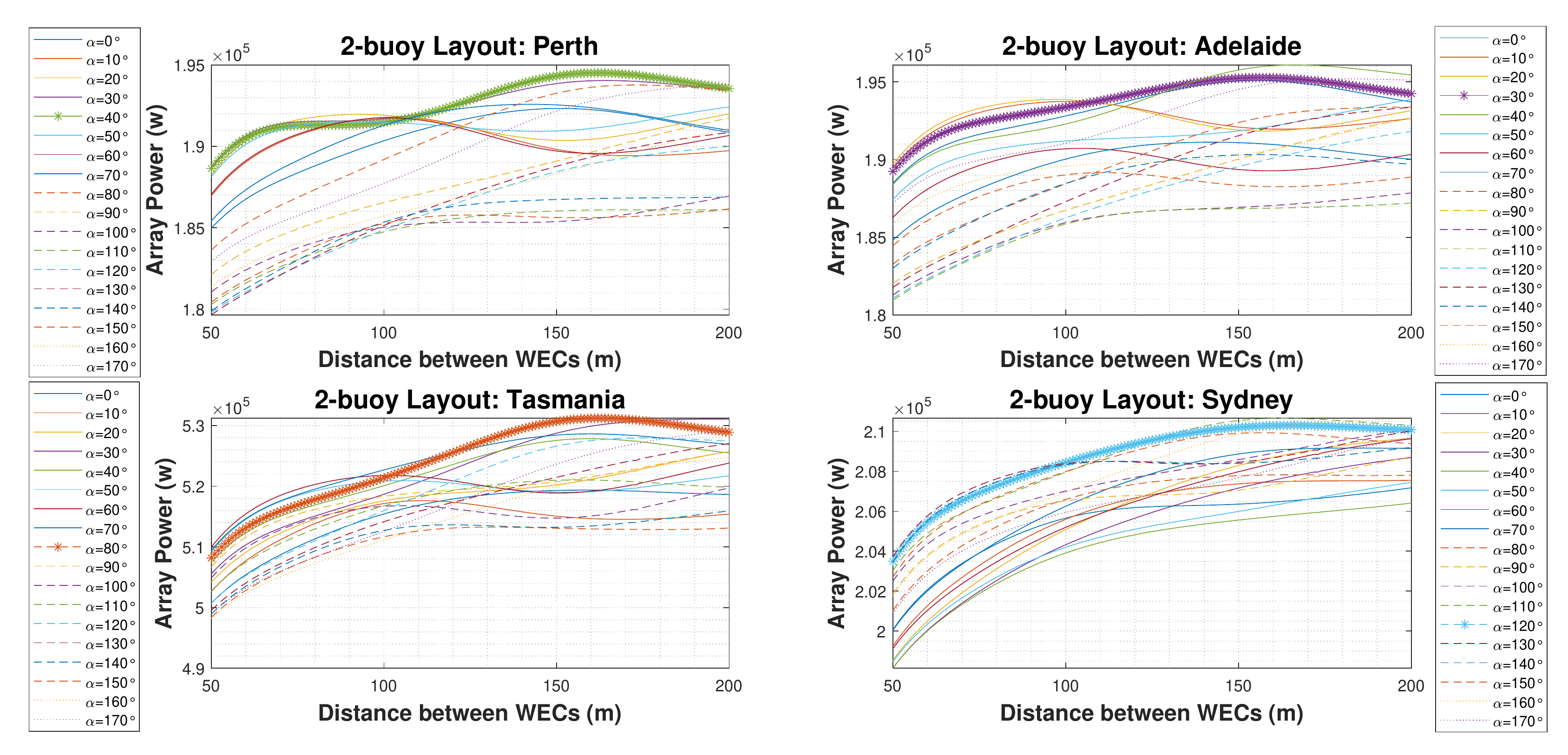}
    \caption{Power output of the two-buoy array over different distances in four wave scenarios.}
    \label{fig:1-2}
\end{figure}

While there are four locations, it can be seen in figure \ref{fig:1-2} that Tasmania has the most wave farm power which is almost 0.534 $Mw$ where the wave direction is 80 degree and distance between WECs is 160 meters. The second location which has considerable farm power is Sydney, although it is 0.21 $Mw$ which is by far less than Tasmania. The power that this layout represents, happened when wave direction is 120 degree and distance is 150 meters. Adelaide and Perth are similar to each other in terms of farm power, range from around 0.18 to 1.96 $Mw$. Both of them indicate good results when wave direction and distance between WECs are 40 degree and 163 meters, respectively. Overall it is clear that the more separation are between converters, the more farm power can be harnessed, except when distance is over 150 meters, mentioned trend does not act as expected.

\subsection{Sensitivity of three-buoy array performance to distance}
\begin{figure}[htb]
    \centering
    \includegraphics[width=\linewidth]{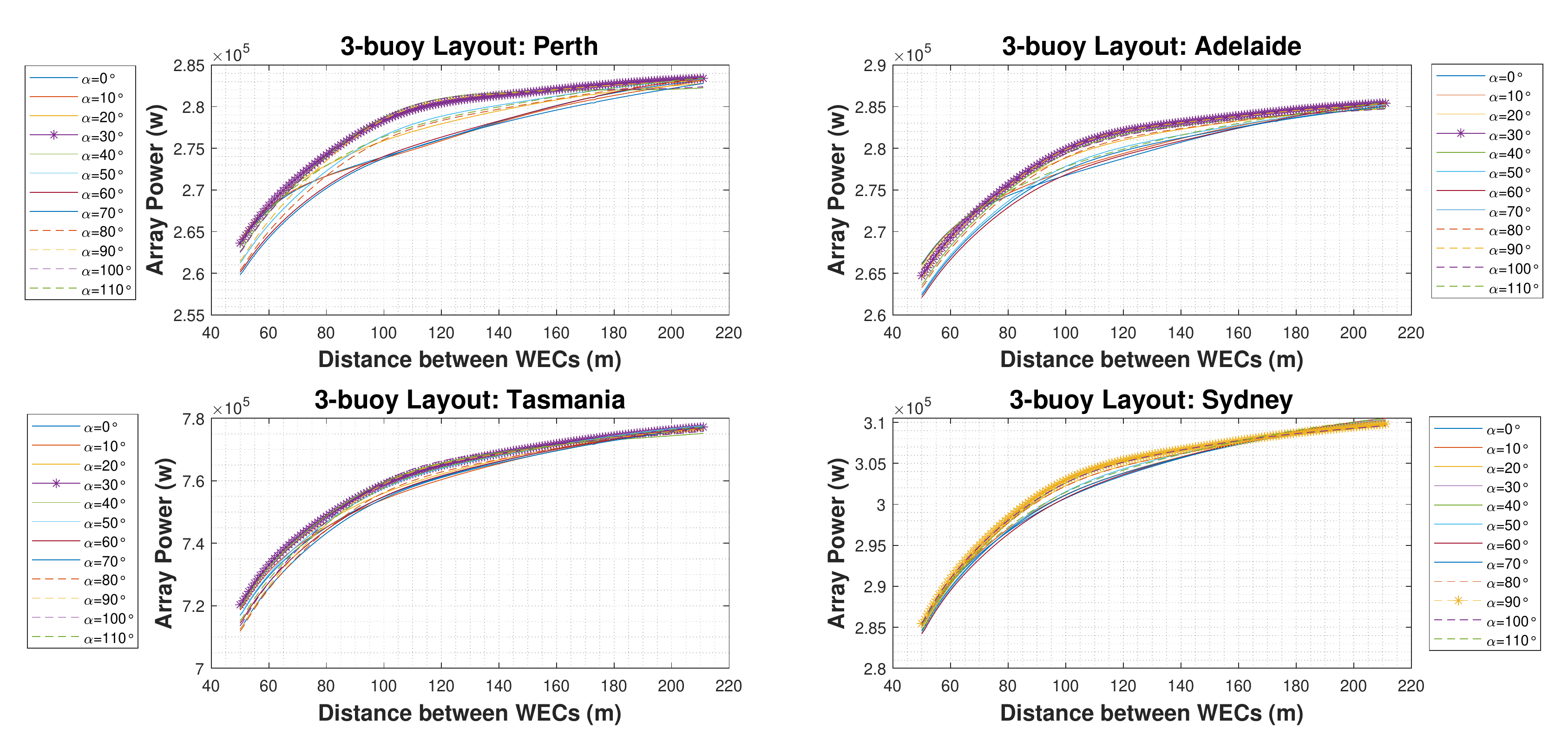}
    \caption{Power output of the three-buoy array over different distances in four wave scenarios.}
    \label{fig:1-3}
\end{figure}

The most common configuration with 3 converters without considering any difference in distances is an equilateral triangle which is used in this research. The most obvious outcome which can be inferred is that the longest distance between WECs, results the highest farm power. The order of places in power of each buoy between locations remained unchanged which would be 0.76, 0.31, 0.286, 0.284 $Mw$ in Tasmania, Sydney, Adelaide and Perth, respectively. Although alpha  is only different in one location, the range of farm power is quite close to each other, Where only 0.05 $Mw$ gap can be witnessed among 12 tested directions. it is evident in figure \ref{fig:1-3} that the wave direction which has the highest farm power on average would be 30 degree in all locations, except Sydney which is 90 degree. It could be mentioned that differences between maximum distance in each layout relates to the limitations that already discussed in section \ref{Section2}.
\subsection{Sensitivity of four-buoy array performance to distance}
\begin{figure}[htb]
    \centering
    \includegraphics[width=\linewidth]{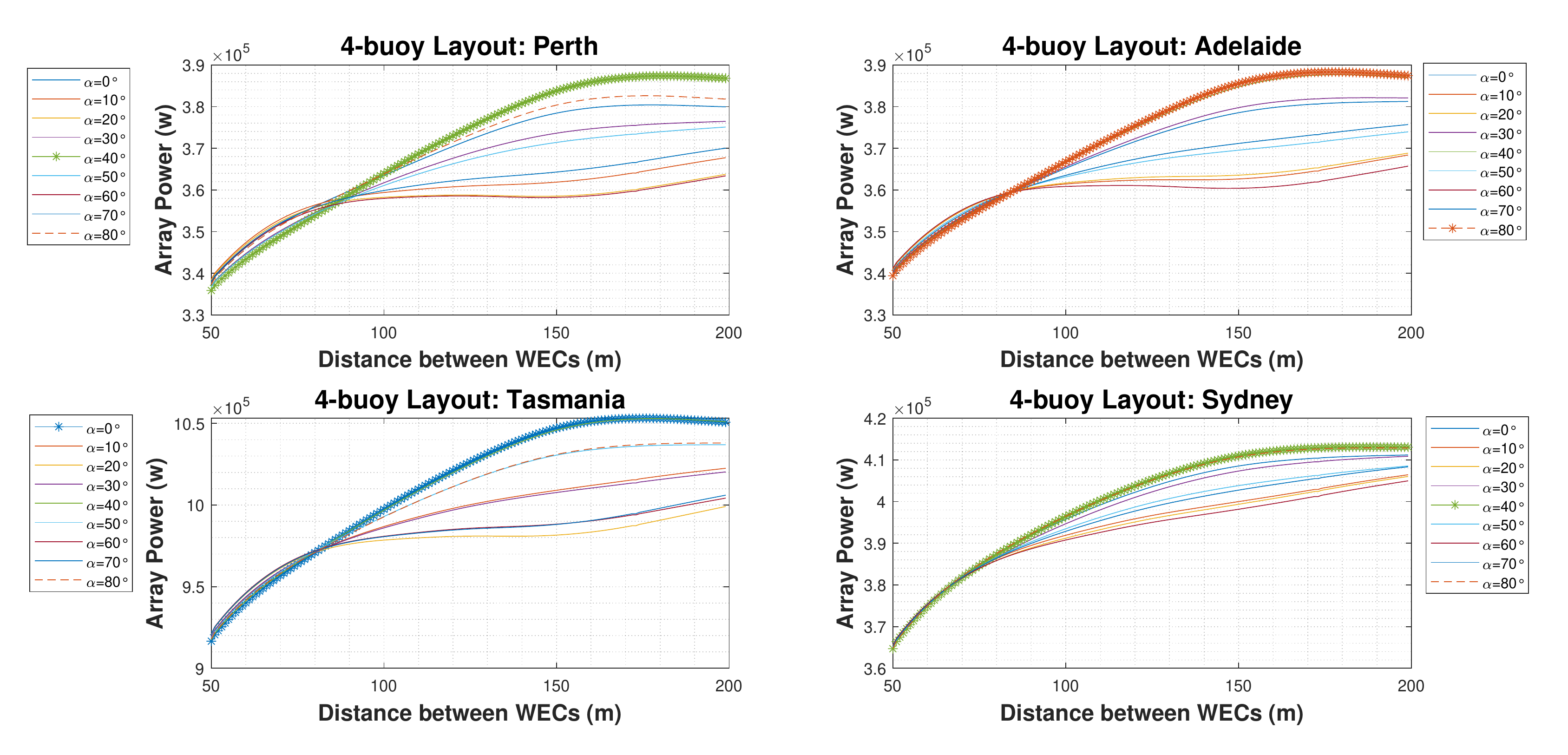}
    \caption{Power output of the four-buoy array over different distances in four wave scenarios.}
    \label{fig:1-4}
\end{figure}

The geometry chosen for four converters is a square shape. Alpha in this layout ranges from 0 to 80 degrees with 10 degree interval. The order of maximum farm power is still the same as mentioned earlier. This measure in Tasmania is 1.05 $Mw$ while the least one is Perth with around 0.387 $Mw$. Turning to the significant wave direction, for Perth, Adelaide and Sydney the direction is 40 degree to harness the maximum farm power, however for Tasmania alpha is 70 degree. Considering the distances between WECs, once again it is apparent that the more separation are between placement of converters, the more wave farm can be extracted. However, at certain distances the outcome leveled and due to this, the minimum distance between WECs is pointed. While in Adelaide and Tasmania 160 and 170 meter distance seems to be the optimal distance between converters, this amount is a bit higher in Sydney and Perth where the highest farm power is firstly seen in 180 meters distance (figure \ref{fig:1-4}).
\subsection{Sensitivity of five-buoy array performance to distance}
\begin{figure}[htb]
    \centering
    \includegraphics[width=\linewidth]{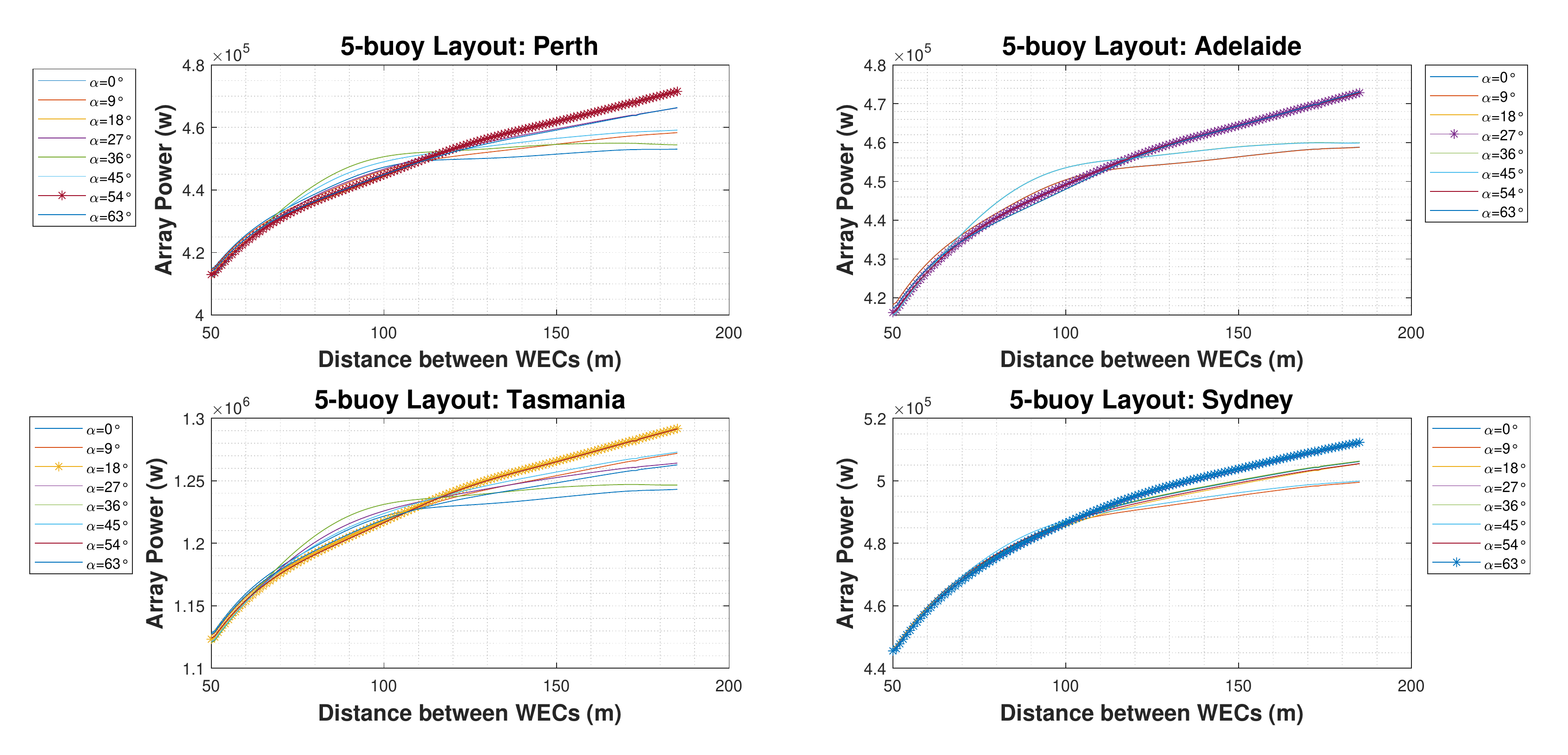}
    \caption{Power output of the five-buoy array over different distances in four wave scenarios.}
    \label{fig:1-5}
\end{figure}

In this study, the configuration of five converters are chosen to be regular pentagon, and because there is no difference between each WECs, the wave direction ranges restricted to be between 0 and 63 with 8 different alphas. The maximum farm power of all four directions is witnessed when wave alpha is either 54 or 63 degree. To be more precise, In Tasmania and Perth optimal alpha is 54 degree, and for the other two it is 63 degree. To consider the distance in this layout, it is obvious that only maximum allowable distance between converters is the optimal choice for this layout (figure \ref{fig:1-5}).
\subsection{Sensitivity analysis of q-factor to the relative angle of rotation}
Interactions need to be measured with a well-known parameter called q factor . A sensitivity analysis has been done by monitoring the q-factor distribution over different rotation angular of the WECs array in different layouts.

\begin{figure}[htb]
    \centering
    \includegraphics[width=0.8\linewidth]{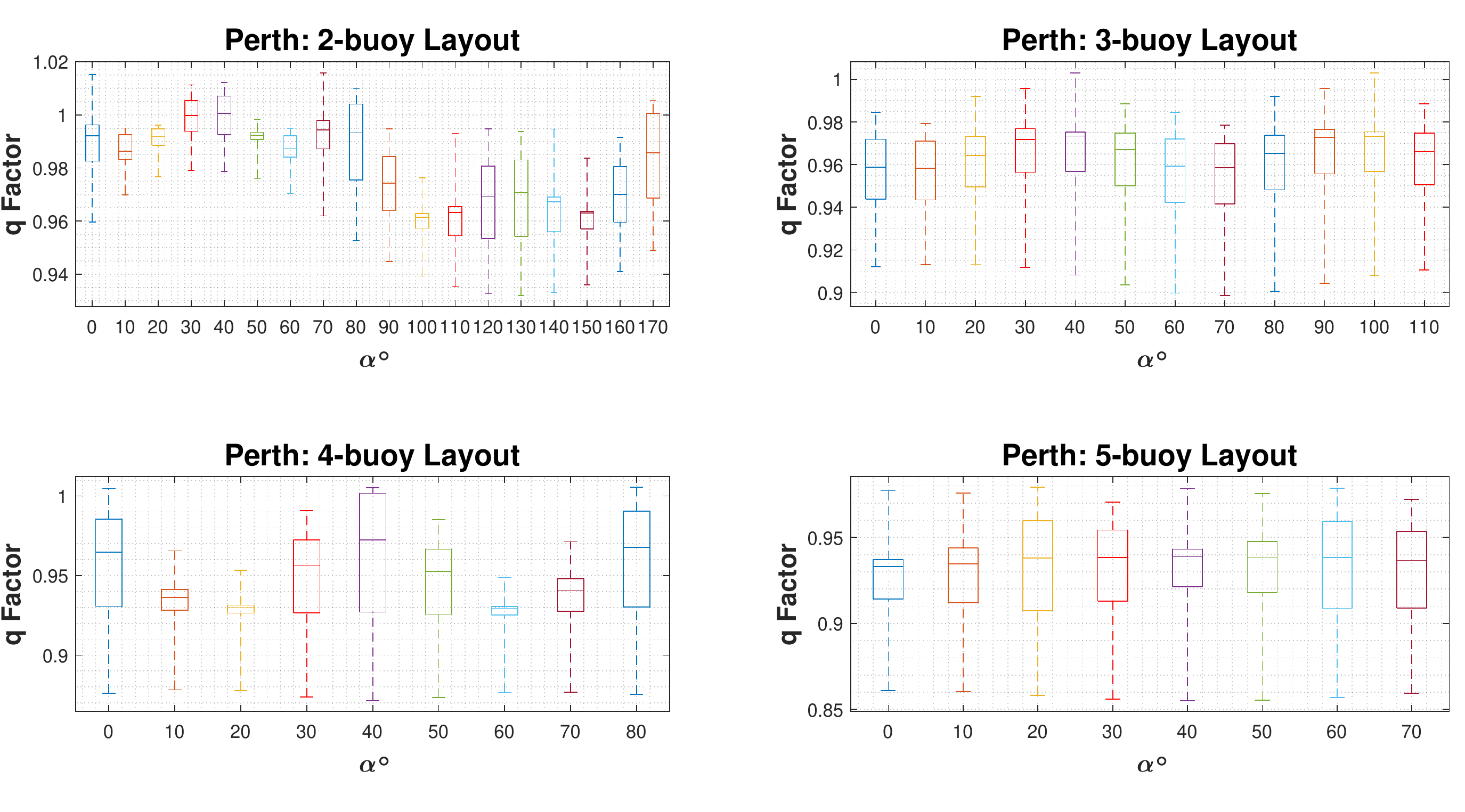}
    \caption{q-factor distribution of the WEC array over different rotation angular due to significant wave direction in Perth wave scenario.}
    \label{fig:2-P}
\end{figure}

\begin{figure}[htb]
    \centering
    \includegraphics[width=0.8\linewidth]{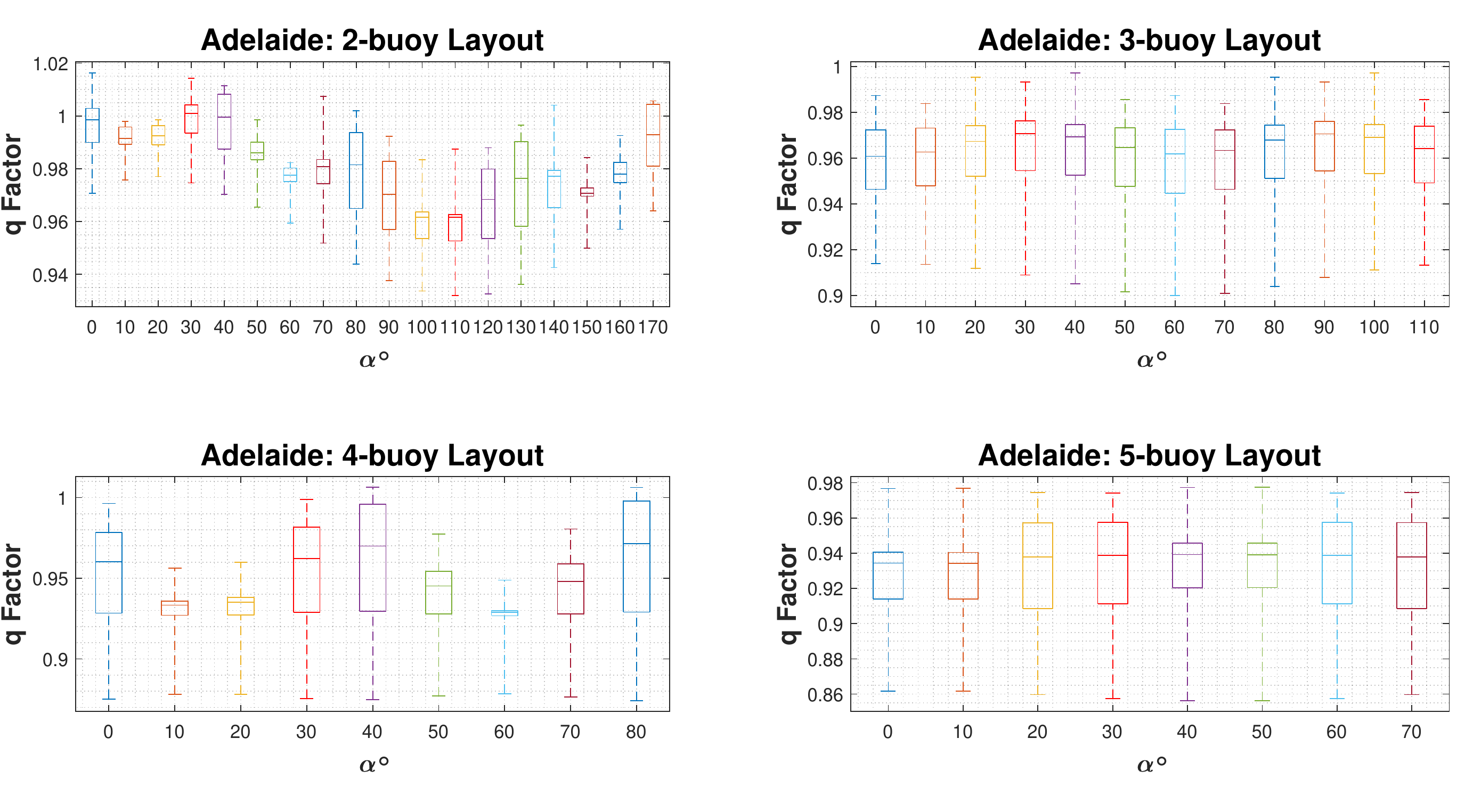}
    \caption{q-factor distribution of the WEC array over different rotation angular due to significant wave direction in Adelaide wave scenario.}
    \label{fig:2-A}
\end{figure}

\begin{figure}[htb]
    \centering
    \includegraphics[width=0.8\linewidth]{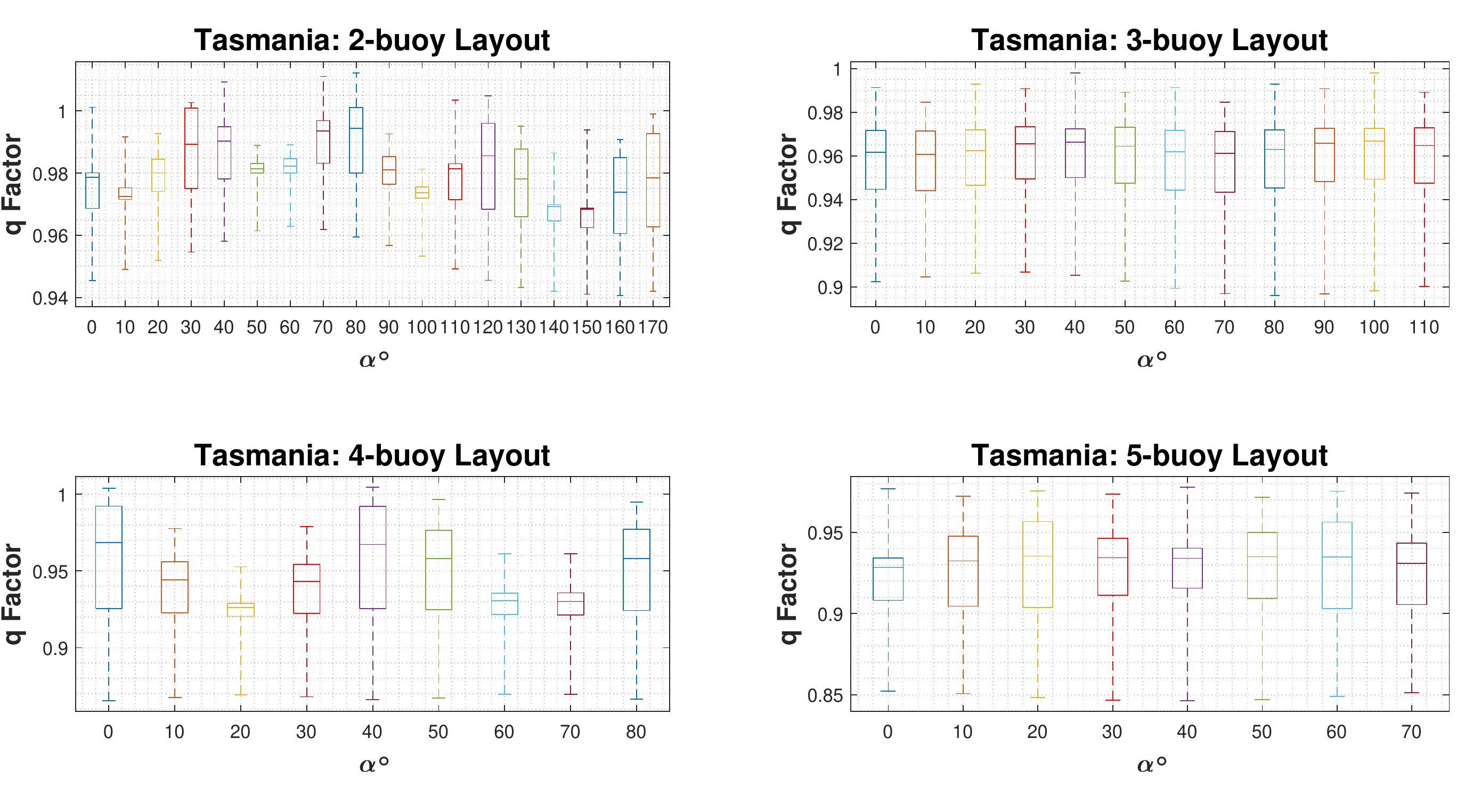}
    \caption{q-factor distribution of the WEC array over different rotation angular due to significant wave direction in Tasmania wave scenario.}
    \label{fig:2-T}
\end{figure}

\begin{figure}[htb]
    \centering
    \includegraphics[width=0.8\linewidth]{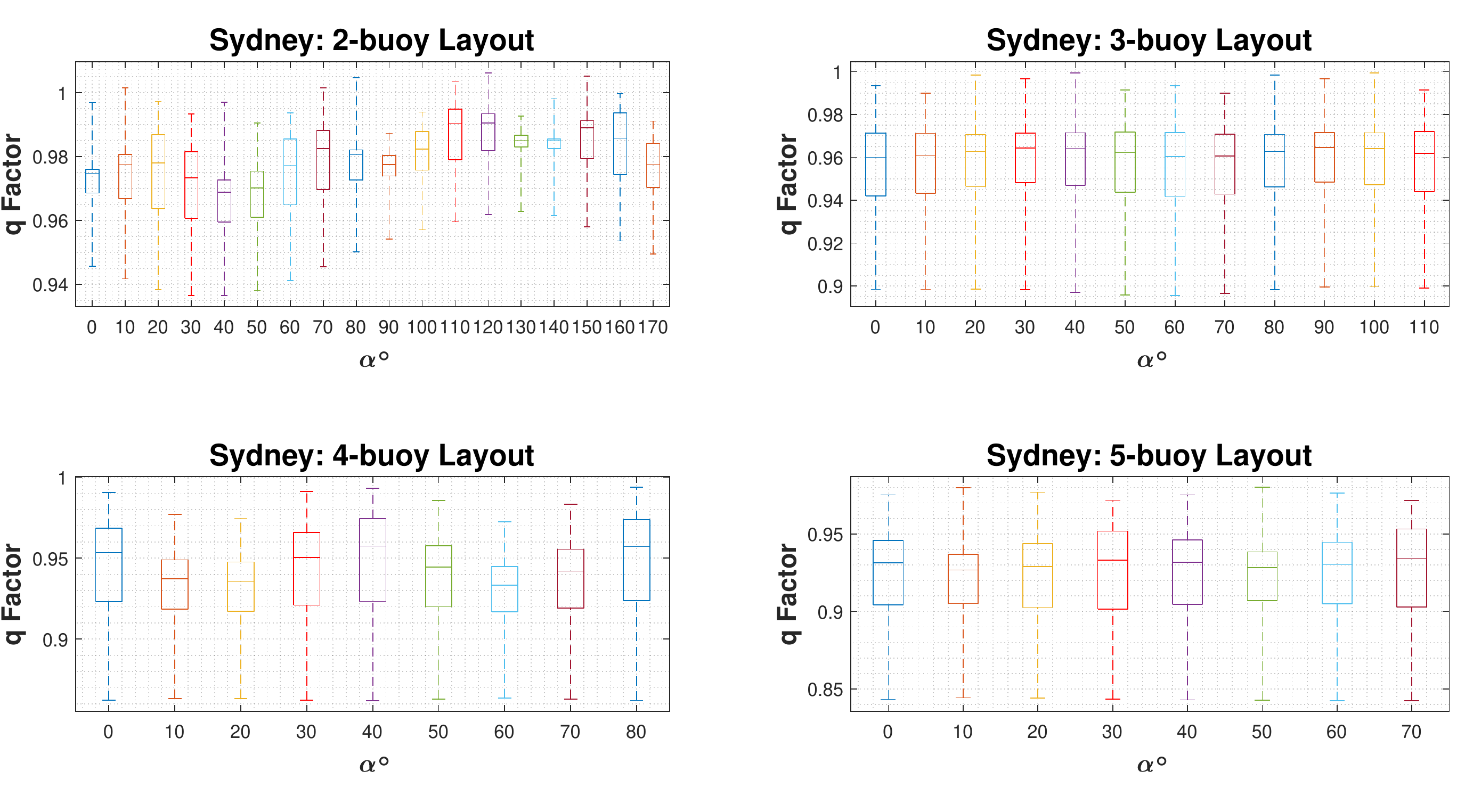}
    \caption{q-factor distribution of the WEC array over different rotation angular due to significant wave direction in Sydeny wave scenario.}
    \label{fig:2-S}
\end{figure}

There is a similarity between locations in Adelaide and Perth in terms of their q factors, but Sydney and Tasmania have seen different trends. In Adelaide and Perth, at first glance at figure \ref{fig:2-A} and figure \ref{fig:2-P} in two-buoy layout, sinusoidal pattern is witnessed and when alpha is between 30 and 40 degrees, the highest q factor is revealed. The distinction between q-factors in three-buoy layout is negligible due to the small effects of significant wave direction on equilateral  triangle buoys, and the maximum q factor can be seen in the 30 and 40 degrees. In four-buoy layout, when significant wave direction is 40 degree, q factor is around 1 in both locations. Among the 9 discussed wave directions, the highest q factors are able to see when alpha is 0, 30, 40 or 80 degree, and average q factor in each directions are a bit over than 0.95. In five-buoy layout, it is clear that average q factors in every directions between all distances are around 0.93, and the range of q factors are from 0.86 to 0.98.
In Tasmania, due to the symmetry between WECs and small effects of changing alpha in q factor for three buoys and five-buoys, changes are not considerable, and the average of q factors in each wave direction is approximately 0.96 and 0.93, respectively. However, in two-buoys when alpha is from 30 to 90 degree, q factors in each distances are between 0.95 and 1.01, while the maximum average of that happens in 80 degree. In four-buoy layout, maximum q factors can be seen in four wave directions which would be 0, 40, 50 and 80 degrees. Further details can be found at in figure \ref{fig:2-T}.
Looking  at Sydney in figure \ref{fig:2-S}, in two-buoy layout, it is evident that maximum q factors are happened when significant wave direction is between 110 and 120 degree. One of the distinctions in comparison to two mentioned locations is the least q factor can be witnessed in 40 degree wave direction, because in order to fulfill one-third of in highest waves all directions, significant wave height is changed. In three-buoy layout the average of q factors in each alpha is around 0.96 which  its changes are not recognizable in all 12 tested alphas. The amount of q factors in four-buoy layout are much similar to the related amounts in Perth and Adelaide where the closest q factors to 1 can be found in 0, 30, 40 and 80 degrees. In five-buoy layout q factors are ranging from 0.84 to 0.98 and the average of this measure is around 0.93 in all 8 wave directions. 

\subsection{Landscape Analysis}
The  figures\label{fig:3-S}to\label{fig:3-T} reveal power for each buoy in four different layouts. Overall it is inferred that the asymmetry in arrays makes the power more predictable in each wave directions and distance.
The illustrated plots in this section indicate four kinds of buoy layout for each areas in Adelaide, Tasmania, Sydney and Perth. As it can be seen in each figure, there are some radial lines in each plot which indicate the buoy locations for different layouts in studied directions with various distances to one another.Indeed, the radial lines are consisted of numerous points which cover the whole area of research. Vertical axis expresses the amount of total power, for each layout, and this amount has been shown in form of a contour.
In general, there is an almost similar pattern for under studied locations only considering the configuration between converters. The maximum amount of extracted energy is more likely to be found in 5-buoy layout. This should be considered that although the power of each converters in some areas for layouts with less number of buoys is considerable, the maximum amounts of extracted energy can be found in even further areas, the asymmetry of these areas is greater, as well. Take 2-buoy layout in Adelaide as an example; various placement angles of buoys and different collision of waves to them, have greater impact on occurrence of asymmetry in such layouts. In some placements of 2-buoy layout, as array experiences different angles, the buoys may see the same significant wave direction. However, in other alphas the shadowing of one buoy to another, can play a crucial role to reduce the energy. This shadowing effect of buoys to each other becomes more drastic, as the number of buoys increase, because in each angle there is more chance of interaction between at least two buoys. 

\begin{figure}[htb]
    \centering
    \includegraphics[width=0.8\linewidth]{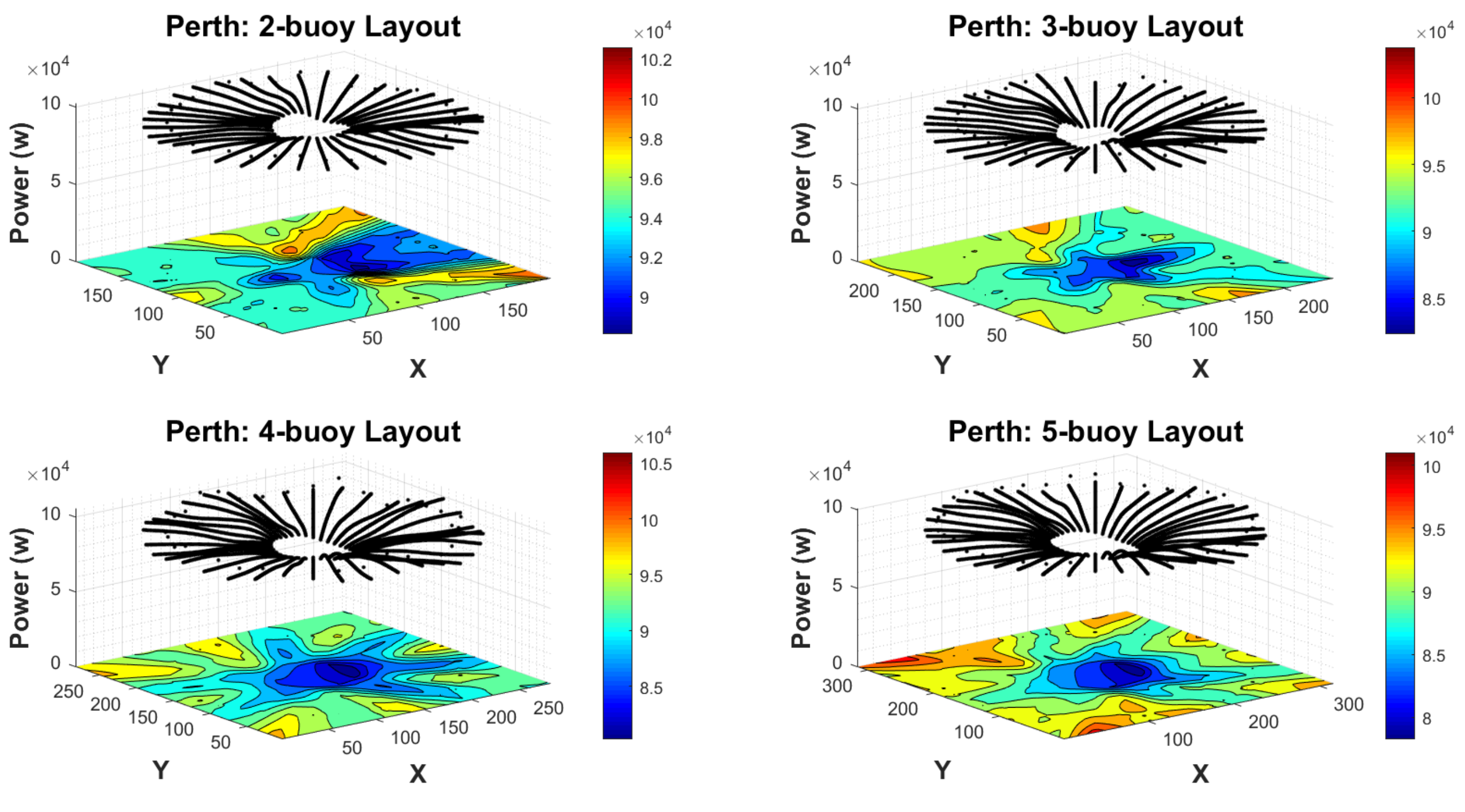}
    \caption{Exploited energy distribution of the WECs array over entire area in Perth wave scenario.}
    \label{fig:3-P}
\end{figure}

\begin{figure}[htb]
    \centering
    \includegraphics[width=0.8\linewidth]{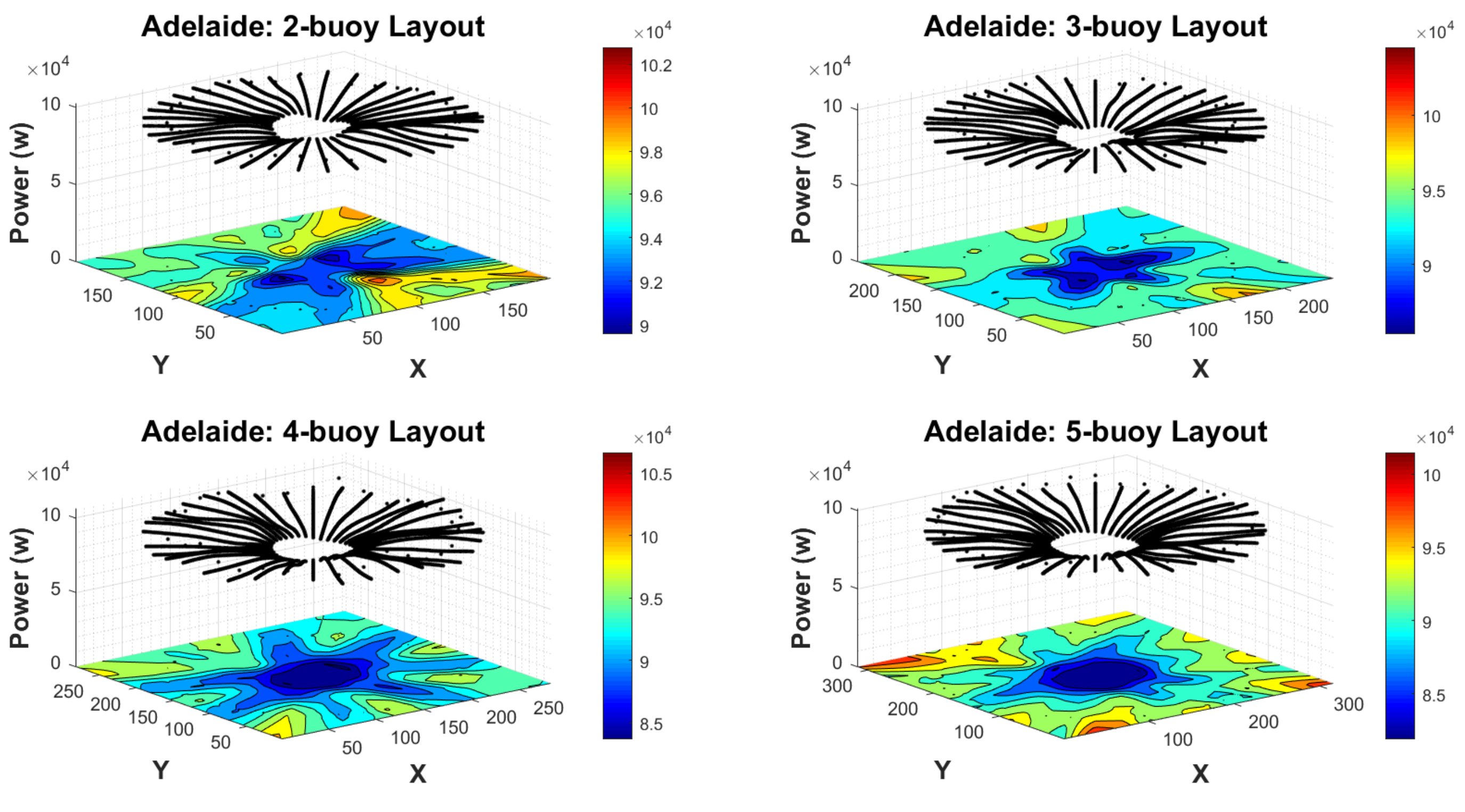}
    \caption{Exploited energy distribution of the WECs array over entire area in Adelaide wave scenario.}
    \label{fig:3-A}
\end{figure}

Taking a look on figure \ref{fig:3-S}, it is interesting that the order of maximum power for each layout is around 0.1 to 0.105 $Mw$ in most areas of the location, however in five-buoy layout this range increases significantly. In this figure ~\ref{fig:3-S}, asymmetry in each layout is more evident than the other locations.
\begin{figure}[htb]
    \centering
    \includegraphics[width=0.8\linewidth]{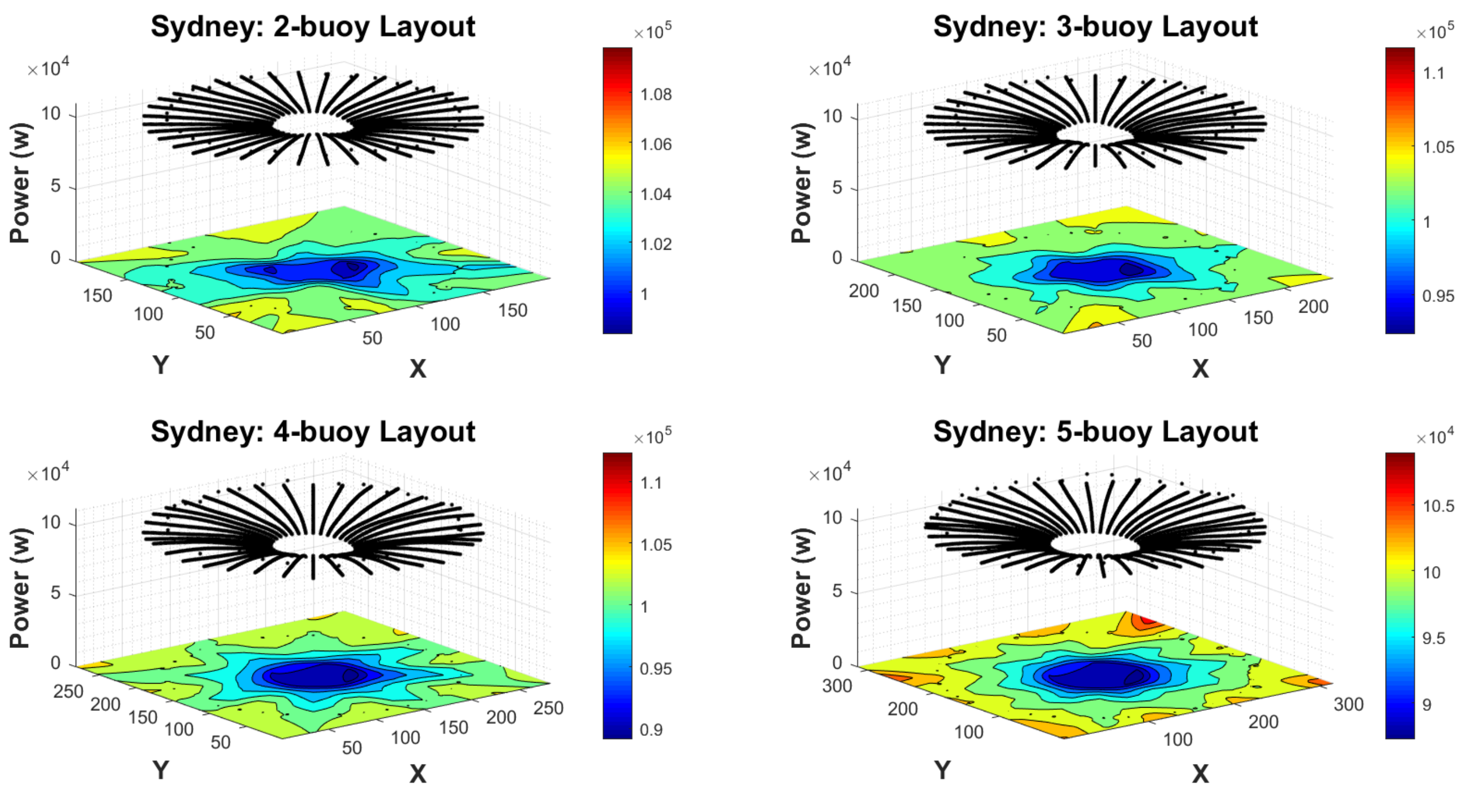}
    \caption{Exploited energy distribution of the WECs array over entire area in Sydeny wave scenario.}
    \label{fig:3-S}
\end{figure}
There are a lot of similarity between Adelaide and Perth in figure \ref{fig:3-P} and figure \ref{fig:3-A}. For instance, in two-buoy layout not only their contour have resemblance but also the range of power is identical, however, based on these plots, more power is extracted in Perth. To compare four-buoy layout, Adelaide is more symmetrical but the chance of reaching 1 $Mw$ power is by far more in Perth.

The highest amount of power can be extracted from converters in Tasmania which is obvious in figure \ref{fig:3-T}. The chance of extracting over than 0.275 $Mw$ power is seen in five-buoy and three-buoy layouts. It is worth to consider 2 buoy layout power when $X$ direction is over than 150 which clearly reveals the potential of this layout under certain condition.
\begin{figure}[htb]
    \centering
    \includegraphics[width=0.8\linewidth]{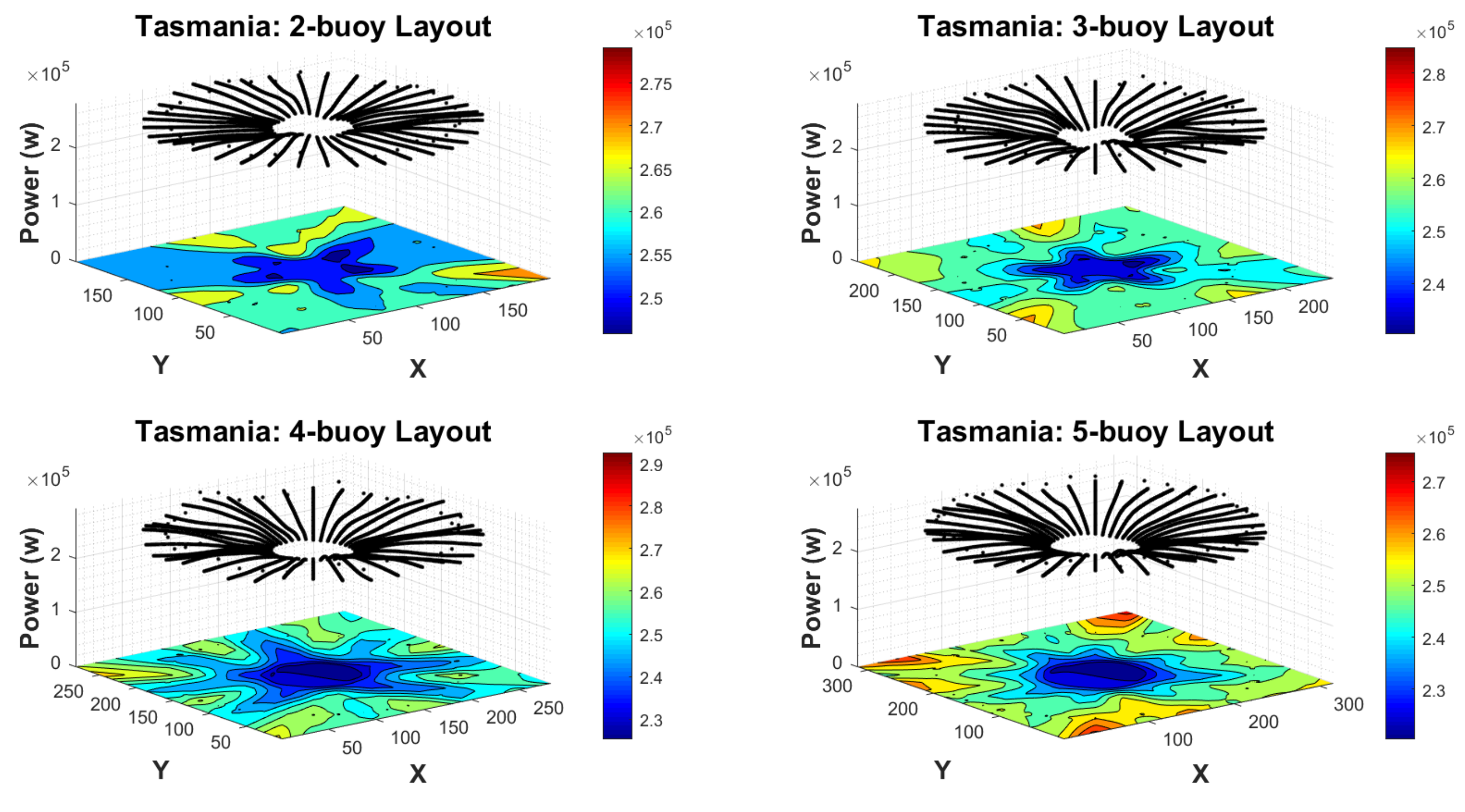}
    \caption{Exploited energy distribution of the WECs array over entire area in Tasmania wave scenario.}
    \label{fig:3-T}
\end{figure}
\subsection{Mean Annual Energy Output Result}
\begin{figure}[htb]
    \centering
    \includegraphics[width=0.8\linewidth]{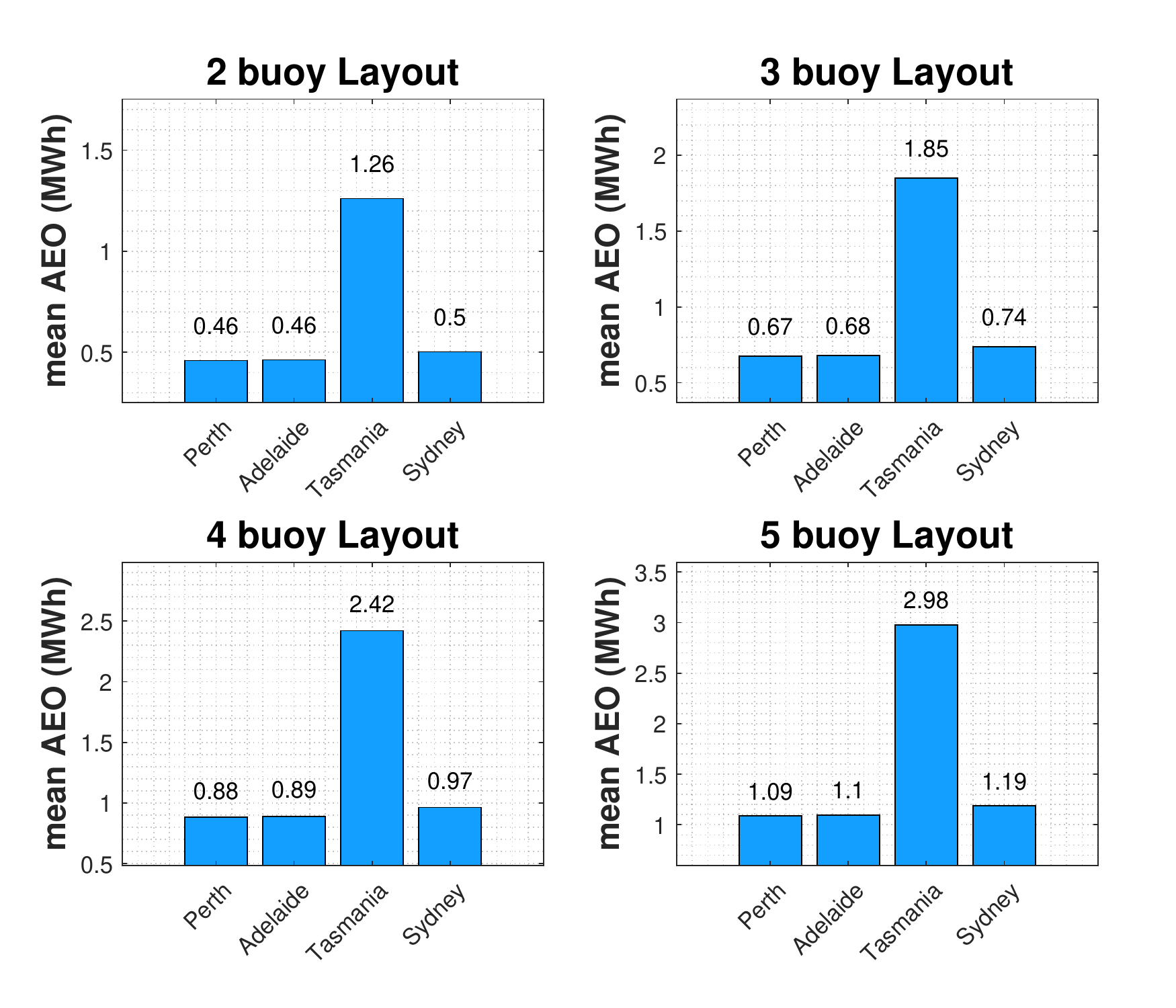}
    \caption{Averaged annual energy output (AEO) as a function of number of buoys in each layout.}
    \label{fig:4}
\end{figure}

The below bar charts reveal the average of energy output annually when the number of buoys increases from 2 to 5 in chosen layout. The most significant facts to emerge from figure \ref{fig:4} are that annual energy output in Tasmania is by far higher than the other locations' AEO, and this parameter has almost the same amount in Perth and Adelaide.
Turning to the Tasmania, it is apparent that 1.26 $Mwh$ can be taken annually by an array with two buoys which is 2.5 times more than Sydney. Not only this proportionality remain intact by increasing the number of buoys to five-buoy layout, but also AEO raises near 0.58 $Mwh$ by adding a converter into the array which would be 1.85, 2.42 and 2.98 for array with 3, 4 and 5-buoy layout, respectively.
Turning to the similarity between Perth and Adelaide, there is a fierce competition among them, however AEO in Adelaide is higher for layouts with more than two buoys. It is interesting to notice that this output in five-buoy layout which is 1.1 $Mw$, is less than AEO in Tasmania for two-buoy layout.

\section {Conclusions}
Wave energy converters can reinforce each other to provide more power output in the form of an array if the distance among them optimises and the arrangement of the layout defines appropriately. In this paper, we analyse the WECs separation in a wave farm with a different number of devices and arrangements. In order to assess the impact of various wave models, we perform and compare all experiments in four real wave scenarios including Sydney, Perth, Adelaide and Tasmania sea sites. According to the experimental analysis, there is a direct relationship between the number of converters and optimal inter-distance among them and also relative angle to significant wave direction. The more separation between converters leads to the more farm harnessed power output. A sensitivity analysis has revealed that the q-factor distribution differed due to different rotation angle of the WECs array. Moreover the mean annual  energy output analysis showed that results in Tasmania is by far higher than the other locations' AEO, and this parameter has almost the same amount in Perth and Adelaide sea sites. Also the landscape analysis approved the maximum amount of extracted energy is more likely to be found in 5-buoy layout.

\section*{Acknowledgements}
The authors would like to appreciate the support of Dr Mehdi Neshat from the University of Adelaide, who was of great help, valuable comments and useful suggestions in this research. And also we would like to appreciate Dr. Nataliia Surgiienko from the University of Adelaide due to publishing the MATLAB source code of the wave energy simulator.  This work has been supported by the High Performance Computing Research Center (HPCRC) - Amirkabir University of Technology under Contract No ISI-DCE-DOD-Cloud-700101-4504.


\bibliographystyle{unsrt}  
\bibliography{sample-bibliography}
\end{document}